\documentclass{PoS}
\usepackage{cite}
\newcommand{\GeV}{${\rm GeV}$}

\title{\vspace*{-3mm}
{\tiny DESY 10--092 \hfill SFB/CPP-10-62}\\
QCD Analysis of the Polarized World Data}

\ShortTitle{QCD Analysis of the Polarized World Data}

\author{\speaker{Johannes Bl\"umlein}\thanks{This work was supported in part by DFG 
Sonderforschungsbereich Transregio 9, Computergest\"utzte
Theoretische Teilchenphysik and the European Commission MRTN HEPTOOLS under Contract No.
MRTN-CT-2006-035505.}\\
        Deutsches Elektronen-Symchrotron, DESY, Platanenallee 6, D-15738 Zeuthen, Germany\\
        E-mail: \email{Johannes.Bluemlein@desy.de}}

\author{Helmut B\"ottcher\\        
Deutsches Elektronen-Symchrotron, DESY, Platanenallee 6, D-15738 Zeuthen, Germany\\
        E-mail: \email{Helmut.Boettcher@desy.de}}

\abstract{
The results of a recent next-to-leading order QCD analysis of the world data on 
polarized deep inelastic scattering are reported. New parameterizations are derived 
for the quark and gluon distributions and the value of $\alpha_s(M_z^2)$ is determined. 
We obtain $\alpha_s^{\rm NLO}(M_Z^2)= 0.1132~~\begin{array}{l}  + 0.0056 \\ -0.0095 
\end{array}$. Potential higher twist contributions to the structure function $g_1(x,Q^2)$ 
are considered.}

\FullConference{XVIII International Workshop on Deep-Inelastic Scattering and Related Subjects, DIS 2010\\
		April 19-23, 2010\\
		Firenze, Italy}

\begin{document}
\sloppy

\section{Introduction}

\vspace*{1mm}\noindent
The short distance behaviour of the partons inside strongly polarized nucleons constitutes one of the
central research topics in QCD being explored both with perturbative and non-perturbative methods.
During the last years the polarized deep-inelastic scattering data have further improved~\cite{
EMCp,E142n,HERMn,E154QCD,E154n,SMCpd,E143pd,HERMpd,E155d,E155p,COMPd,JLABn,CLA1pd,CLA2pd,COMP1}.
In this note we report on a new next-to-leading order (NLO) QCD analysis of these data~\cite{BB10}, 
updating earlier investigations~\cite{BB02}. At large enough four-momentum transfer $Q^2 = -q^2$, the 
structure function $g_1(x,Q^2)$ mainly receives twist--2 contributions\footnote{Twist--3 contributions are 
connected by target mass effects, cf.~\cite{BT}.} and is related to the polarized twist--2 parton 
distribution functions (PDFs). We analyze the structure function $g_1(x,Q^2)$, which is derived from 
the longitudinal 
polarization asymmetry accounting for a data-based description of the denominator 
function~\cite{F2NMC} and corresponding parameterizations for the longitudinal structure function, cf. 
\cite{BB10}. In the present analysis we include the $O(\alpha_s)$ contribution due to charm quarks,
\cite{HEAV1,BRN}~\footnote{The $O(\alpha_s^2)$ heavy flavor corrections are only known in the 
asymptotic region $Q^2 \gg m^2$, \cite{HQ2}.}. The structure function $g_2(x,Q^2)$ is described at 
leading twist by the Wandzura-Wilczek relation \cite{WW,BRN}. The 
parameters of the polarized parton densities, which can be measured using the above data sets, are 
determined with correlated errors along with the QCD scale $\Lambda_{\rm QCD}^{N_f = 4}$.
We also analyze potential contributions of higher twist. Finally, a phenomenological parameterization 
of the polarized NLO parton distribution functions is provided in terms of  grids for the central 
values and correlated errors, \cite{GRID}.
\section{The Analysis}

\vspace*{1mm}\noindent
The NLO QCD analysis of the structure function $g_1(x,Q^2)$ is performed in Mellin space following
the standard formalism, cf. e.g.~\cite{BV}, including the heavy quark corrections 
\cite{AB1}.~\footnote{We refrain from carrying out small-$x$ resummations, since yet 
unknown subleading terms are very likely to cancel the leading order effects, cf. \cite{BV1}.}  In 
this 
representation the evolution equations can be solved analytically, in both a fast and 
numerically precise way. Only one numerical integral around the singularities of the solution in 
the complex plane, located on the real axis left to an upper bound, has to be performed to represent
$g_1(x,Q^2)$. As the data are located at low values of $Q^2$ target mass corrections are applied, 
cf.~\cite{BT,BT1}. For the deuteron targets a wave function correction is performed~\cite{OMEGD}.
The parton distributions at the starting scale $Q^2_0 = 4~\GeV^2$ are parameterized by
\begin{eqnarray}
\label{eq1}
x \Delta f_i(x,Q_0^2) =  \eta_i A_i x^{a_i} (1-x)^{b_i} (1+\gamma_i x)~,
\end{eqnarray}
with $\eta_i$ the first moments. The present analysis parameterizes the sea quarks assuming 
approximate flavor $SU(3)$ symmetry. The deep-inelastic data alone cannot resolve the flavor 
dependence of the sea. Taking into account semi-inclusive data \cite{HERMES1}, and later on polarized 
Drell-Yan and di-muon data, will allow the determination of polarized sea quark distributions, 
similar to the unpolarized case.~\footnote{For a first analysis accounting for the flavor 
dependence of the sea quarks see \cite{DSSV}.} $\eta_{u_v}$ and $\eta_{d_v}$ are fixed due to 
the neutron and hyperon-$\beta$ decay parameters $F$ and $D$, which are very well measured~:
\begin{eqnarray}
\eta_{u_v} - \eta_{d_v} = F+D &{\rm and}&  \eta_{u_v} + \eta_{d_v} = 3F + D~, \\
\eta_{u_v} = 0.928 \pm 0.014   &{\rm and}&  \eta_{d_v} = -0.342 \pm 0.018~. 
\end{eqnarray}
The parameters in (\ref{eq1}) cannot all be measured using the present data since for some the
$\chi^2$--fit yields errors larger than 100\%. In case of the sea-quark and gluon density $\gamma_i$
is found to be compatible with zero. Furthermore the $a_i$-parameters of the distributions $\Delta 
q_s$ 
and $\Delta G$
are related by about $a_{\Delta G} = a_{\Delta q_s} + 1$, which we use. The parameters $\gamma_{u_v}$ 
and $\gamma_{u_v}$ are fitted in an intial run and are then kept fixed as model parameters. For the 
large-$x$ parameters $b_{\Delta q_s}$ and $b_{\Delta G}$ we used the relation $b_{\Delta 
q_s}/b_{\Delta G}$(pol) = $b_{\Delta q_s}/b_{\Delta G}$(unpol) = 1.44 and determine $b_{\Delta G} = 
5.61$ and $b_{\Delta q_s} = 8.08$ in the fit. In the final fit 8 parameters are determined including
$\Lambda_{\rm QCD}^{N_f = 4}$. In Figure~1 we show the four distributions $\Delta u_v, 
\Delta d_v, \Delta q_s$ and $\Delta G$ at the input scale and compare them to other determinations.
In Ref.~\cite{BB10} we 

\begin{figure}[htb]
\begin{center}
\includegraphics[angle=0, width=10.0cm]{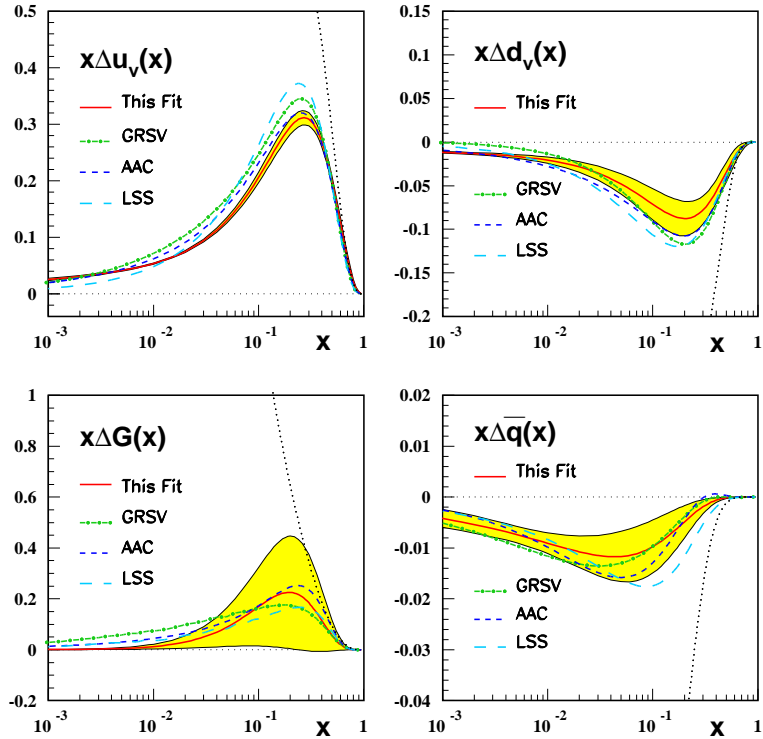}
\end{center}
\caption{\label{fig:xpdf}
\small
NLO polarized parton distributions at the input scale
$Q_0^2 = 4.0~\GeV^2$ (solid line) compared to results obtained by
GRSV~(dashed--dotted line)~\cite{GRSV}, AAC~(dashed line)~\cite{AAC}, 
and LSS~(long dashed line)~\cite{LSS}.
The shaded areas represent the fully correlated $1\sigma$ error bands
calculated by Gaussian error propagation. The dotted line indicates the positivity bound
using the parameterization \cite{MSTW}; from Ref.~\cite{BB10}.}
\end{figure}
\noindent
provide the correlated errors. Due to this one may perform Gaussian error 
propagation for all observales based on polarized parton denstities predicting the PDF 
errors of these quantites. There we also compute a series of moments for the different parton 
densities which can be compared to upcoming lattice simulations. 

\begin{figure}[htb]
\begin{center}
\includegraphics[angle=0, width=9.0cm]{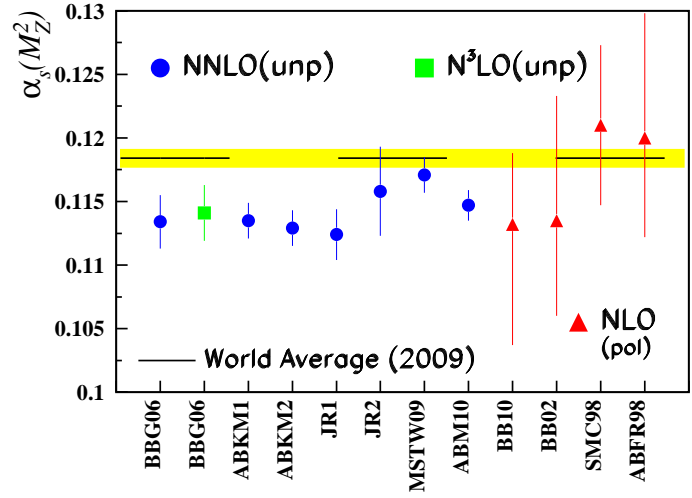}
\end{center}
\caption{\label{fig:alp}
{\small
A summary of the current measurements of $\alpha_s(M_Z^2)$ from unpolarized and polarized
DIS data, cf. Ref.~\cite{BB10} for details. Due to the size of errors we include only the
results of NNLO and N$^3$LO analyses in the unpolarized case, while those in the polarized case
stem from NLO analyses. The yellow band marks the world average of 2009 \cite{BETHKE}.
\vspace*{-4mm}
}} 
\end{figure}

The nucleon spin is given by the relation
\begin{eqnarray}
\frac{1}{2} = \frac{1}{2} \langle \Delta \Sigma(x) \rangle_0 + \langle \Delta G(x) \rangle_0
+ L_q + L_g
\end{eqnarray}
to the first moments of the polarized flavor singlet and gluon distributions and the quark and gluon 
angular momenta $L_{q,g}$. In the present analysis we obtain
\begin{eqnarray}
\langle \Delta \Sigma(x) \rangle_0  &=& 0.216 \pm 0.079 \\
\langle \Delta G(x) \rangle_0       &=& 0.462 \pm 0.430~, 
\end{eqnarray}
which saturates the required value even for vanishing values for $L_{q}$ and $L_{g}$. However, the
error on the gluon density is still rather large.

In deep-inelastic QCD analyses it is important to determine the PDF-parameters at the initial scale
$Q_0^2$ {\it together} with the QCD scale $\Lambda_{\rm QCD}$ since there are strong correlations,
e.g. between the gluon-normalization and $\alpha_s(M_Z^2)$, but also to other parameters. We obtain
\begin{eqnarray}
\Lambda_{\rm QCD}^{(4)} = 243 \pm 62~~(\rm exp)~~\begin{array}{l} -37 \\ +21 \end{array}~{\rm (FS)}
~~\begin{array}{l} +46 \\ -87 \end{array}~{\rm (RS)}~~~{\rm MeV}~.
\end{eqnarray}
The renormalization (RS) and factorization scales (FS) were varied by a factor of 2 around $Q^2$.
Here we excluded values $\mu_{f,r}^2 < 1~\GeV^2$, unlike in Ref.~\cite{BB02}, since at
scales lower than 1~$\GeV^2$ the perturbative description cannot be considered reliable
anymore. Correspondingly, one obtains 
\begin{eqnarray}
\alpha_s(M_Z^2)  = 0.1132
~~\begin{array}{l}  + 0.0043 \\ -0.0051 \end{array}~{\rm (exp)}
~~\begin{array}{l}  - 0.0029 \\ +0.0015 \end{array}~{\rm (FS)}
~~\begin{array}{l}  + 0.0032 \\ -0.0075 \end{array}~{\rm (RS)}~.
\end{eqnarray}
The errors are much larger than in the unpolarized case, at NNLO, where an accuracy of $O(1\%)$ is 
reached.
Still the central value is lower  than the current world average and well comparable to the 
unpolarized values. In Figure~2 we summarize the current status of $\alpha_s(M_Z^2)$ measurements 
in deep-inelastic scattering.
We would like to mention the results of the unpolarized NS-analysis \cite{BBG} at N$^3$LO~:

\vspace*{-7mm}
\begin{eqnarray}
\alpha_s(M_Z^2)  = 0.1141 {
                           \begin{array}{l} + 0.0020 \\
                                            - 0.0022 \end{array}}  
\end{eqnarray}
and recent combined NS and singlet  NNLO analyses \cite{JR,ABKM}
\begin{eqnarray}
\alpha_s(M_Z^2)  &=& 0.1124 \pm 0.0020 \\
\alpha_s(M_Z^2)  &=& 0.1135 \pm 0.0014~. 
\end{eqnarray}
Very recently, due to the inclusion of the combined H1+ZEUS data the latter value receives a
slight change to   
\vspace*{-3mm}
\begin{eqnarray}
\alpha_s(M_Z^2)  = 0.1147 \pm 0.0012~,
\end{eqnarray}
reaching now the accuracy of 1~\%, \cite{ABM}.

We also fitted additive higher twist terms to $g_1(x,Q^2)$ to explore the
corresponding structures in the region $x \leq 0.6$ for the proton- and deuteron targets.
While in case of the deuteron target the results is fully compatible with zero, an effect   
of up to 2~$\sigma$ is observed for four of five bins in case of the proton target. This result is 
indicative mainly.
In the unpolarized case one has much better means to separate leading and higher twist effects, 
cf.~\cite{BB1,SA2}, in a clear manner. This also requires even higher order corrections. A comparable 
analysis in the polarized case has to be based on much more precise data 
in a far wider range of $Q^2$ which can be obtained at future colliders such as the EIC.


\end{document}